# Finite momentum Cooper pairing in 3D topological insulator Josephson junctions

Angela Q. Chen[1,†], Moon Jip Park[1,†], Stephen T. Gill[1], Yiran Xiao[1], Gregory J. MacDougall[1], Matthew J. Gilbert[2], Nadya Mason[1,*]

[1]Department of Physics and Frederick Seitz Materials Research Laboratory, University of Illinois, Urbana, IL, USA

[2]Department of Electrical and Computer Engineering, University of Illinois, Urbana, IL, USA

[*]Correspondence to: nadya@illinois.edu

[†]A. Q. Chen and M. J. Park contributed equally to this work.

## Abstract

Unconventional superconductivity arising from the interplay between strong spin-orbit coupling and magnetism is an intensive area of research. One form of unconventional superconductivity arises when Cooper pairs subjected to a magnetic exchange coupling acquire a finite momentum. Here, we report on a signature of finite momentum Cooper pairing in the 3D topological insulator $Bi_2Se_3$. We apply in-plane and out-of-plane magnetic fields to proximity-coupled $Bi_2Se_3$ and find that the in-plane field creates a spatially oscillating superconducting order parameter in the junction as evidenced by the emergence of an anomalous Fraunhofer pattern. We describe how the anomalous Fraunhofer patterns evolve for different device parameters, and we use this to understand the microscopic origin of the oscillating order parameter. The agreement between the experimental data and simulations shows that the finite momentum pairing originates from the coexistence of the Zeeman effect and Aharonov-Bohm flux.

## Introduction

In the conventional Bardeen-Cooper-Schrieffer (BCS) theory of superconductivity, Cooper pairs form an isotropic condensate with a zero center-of-mass momentum[1]. However, introducing magnetism can change the stability of the BCS superconducting state, thereby destroying superconductivity or, as in unconventional superconductors, altering the pairing symmetry[2]. The potential for unconventional superconductivity at the confluence of magnetism and superconductivity has made it an area of great theoretical and experimental interest. One such example of an unconventional superconducting state is Fulde-Ferrell-Larkin-Ovchinnikov (FFLO) superconductivity, which was proposed as a way for maintaining superconductivity even beyond the critical Zeeman field [3,4]. Despite the intensive search for an FFLO superconductor in various types of materials such as heavy fermion compound $CeCoIn_5$ [5,6] and BEDTTTF-based organic superconductors[7,8,9], FFLO superconductivity still remains a controversial subject[2,10,11,12].

To better hunt for unconventional superconductivity, there have been proposals for utilizing materials with strong spin-orbit interaction coupled to a conventional s-wave superconductor. This is predicted to stabilize an FFLO superconducting state: the spin-orbit coupling lifts the degeneracy in the Fermi surfaces of the material and introduces an anisotropy to the surfaces that makes it more amenable to a finite momentum phase[13,14,15]. In particular, time-reversal invariant topological insulators (TIs), whose surface states are massless Dirac fermions, are proposed to be an attractive candidate for unconventional superconductivity that carries finite momentum pairing[16]. To the best of our knowledge, experimental signatures of finite momentum Cooper pairs in TIs have mainly been sought after in the electron-doped regime of the 2D TI HgTe quantum wells[17], but the surface states of 3D TIs also provide unique advantages to engineering finite momentum pairing. The Dirac cones on the surfaces are non-degenerate and have spin-momentum locking. As a consequence, the Fermi surface of the Dirac cone shifts uni-directionally under the application of an in-plane magnetic field to the surface, which can lead to an FFLO state[13,16]. Even though transport measurements in normal 3D TIs are often complicated by the presence of bulk carriers, there is experimental consensus that the metallic surface state dominates transport in a proximity-coupled TI even when the bulk is not depleted[18,19,20].

To this end, we study the experimental signatures of Cooper pairs in a superconductor (S)-3D TI-S Josephson junction subjected to in-plane and out-of-plane magnetic fields. We probe the phase of Cooper pairs by generating Fraunhofer patterns with an out-of-plane field, and we find that adding an in-plane field distorts the Fraunhofer patterns by (1) transferring the intensity of superconductivity from the central Fraunhofer peak at $B_z = 0$ out to finite magnetic field values and (2) introducing asymmetries between positive and negative values of $B_z$ in the Fraunhofer patterns. We show that the intensity transfer is suggestive of a spatially oscillating superconducting order parameter phase and that the asymmetry is a function of sample geometry. Simulations show a close match between experimental data and finite momentum Cooper pair theory.

## Results

### Experimental set-up

Our devices consist of $Bi_2Se_3$ flakes that are mechanically exfoliated onto $Si/SiO_2$ substrates and contacted with two superconducting electrodes, forming a Josephson junction device. The measured devices vary in flake thicknesses and junction dimensions (Table 1). A representative atomic force microscope image is shown in Fig. 1a for device 1, which has flake thickness $t \sim 9$ nm, average junction width $W \sim 860$ nm, and electrode spacing of 140 nm. Because we utilize high in-plane fields to tune the behavior of the junction, we choose NbTi/NbTiN ($T_c \sim 12.5$ K and $H_{c2,\text{in-plane}} > 9$ T) as the superconducting material.

Out-of-plane magnetic field $B_z$ is applied to the superconducting junction to generate a Fraunhofer pattern. In the absence of any in-plane fields, the devices exhibit a central peak with maximum critical current $I_c$ and decaying side peaks, as is expected for a conventional Fraunhofer pattern. The nodes of our Fraunhofer pattern are at $B_z = \frac{n\Phi_0}{Wd}$, where magnetic flux quantum $\Phi_0 = \frac{h}{2e}$, $d$ is the effective electrode spacing that takes into account flux focusing (see supplementary material), and $n$ is an integer[21]. The Fraunhofer pattern of device 1, shown in Fig. 1b, is representative of our devices.

When an in-plane field along the current direction $B_y$ is introduced to the devices, the conventional Fraunhofer pattern is modulated to an anomalous Fraunhofer pattern: the Fraunhofer pattern is shifted along the $B_z$ direction and the critical current of the side lobes increases as the critical current of the central lobe disappears. To measure the evolution of the Fraunhofer pattern as a function of $B_y$, we apply a small AC excitation (with zero DC current) and measure the differential resistance $dV/dI$ as a function of $B_z$ and $B_y$, where lower resistance corresponds to higher critical current, similar to in Ref. 17. The evolution of the Fraunhofer pattern for device 1 is shown in Fig.

2a. As $B_y$ is applied, the Fraunhofer pattern is shifted along $B_z$. This is evident as an overall tilt of the 2D differential resistance map in Fig. 2a. Although a tilt could be caused by misalignment of the sample with respect to the $B_y$-$B_z$ plane, this type of misalignment would cause a similar shift in the Fraunhofer pattern when a field is applied in any in-plane direction. Because we do not see a corresponding shift when an in-plane field perpendicular to current direction $B_x$ is applied, we can exclude sample misalignment as a cause for the shift of the Fraunhofer pattern.

Besides the shift to the Fraunhofer pattern, we also observe additional side branch features in the evolution of the Fraunhofer pattern: at $B_y = 0$, the intensity of superconductivity is maximum at the central lobe, but as $B_y$ is increased, the intensity is transferred outwards to higher values of $B_z$. The emergence of this anomalous side branch becomes more evident if the tilt in the 2D differential resistance map is removed by rotating the graph until the lobe minima are vertical, as shown in Fig. 2b. The approximate locations of the minima of the lifted side lobes are marked as a guide to the eye, and a slope for the side branch can be approximated, as illustrated by the dashed black line in Fig. 2b (see supplementary material). This unique Fraunhofer evolution is evidence of finite momentum pairing, as discussed below.

**Modeling Josephson junction with finite momentum pairing**

To determine the origin of the evolution of the Fraunhofer pattern, we begin by considering the mesoscopic effects of magnetic fields $B_z$ and $B_y$ on the superconducting order parameter phase. $B_z$ generates a spatially modulated phase difference along the $\hat{x}$-direction between the two superconducting leads. The phase modulation manifests itself as a spatially oscillating current distribution in the $\hat{x}$-direction ($I(x) = i_c sin(\Delta\phi - \frac{2eB_z xd}{c})$) [21]. By summing up all the oscillating components of the current, the conventional Fraunhofer diffraction pattern arises, which can be derived from the following equation:

$$I_c = i_c W \left| \frac{\sin\left(\frac{\pi\Phi}{\Phi_0}\right)}{\frac{\pi\Phi}{\Phi_0}} \right|. \qquad (1)$$

While the Fraunhofer pattern is generated by $B_z$, the additional high in-plane magnetic field, $B_y$, generates a Zeeman effect within the bands and adds magnetic flux inside the topological insulator flake. When the in-plane Zeeman effect is present, the low energy Hamiltonian of the TI surface can be written as

$$H_{Dirac} = -\hbar v_F \left(k_x - \frac{g\mu B_y}{\hbar v_f}\right)\sigma_y + \hbar v_f k_y \sigma_x, \qquad (2)$$

where $v_F$ is the Fermi velocity of the Dirac cone, $g$ is the $g$-factor, and $\mu$ is the Bohr magneton[22]. By examining the Hamiltonian, we find that the location of the Dirac node is shifted from the Γ-point along the $\hat{x}$-direction by $\frac{g\mu B_y}{\hbar v_f}$, resulting in a shift of the corresponding Fermi surface. Fig. 3a shows a schematic of the shifted Fermi surface. As a result of this shift, when the electrons on the TI Fermi surface form spin singlet Cooper pairs, the Cooper pairs gain a finite center of mass momentum $\frac{2g\mu B_y}{\hbar v_f}$. As a consequence, the superconducting order parameter at the end of each junction has a phase modulation in the $\hat{x}$-direction. The order parameter is given as $\Delta_{L,R} \approx \Delta_0 e^{i\frac{2g\mu B_y}{\hbar v_f}x}$, where $\frac{2B_y x g\mu}{\hbar v_f}$ is the phase modulation due to a finite momentum shift. The finite momentum of the Cooper pair under these conditions is similar in nature to FFLO states[2,3]. Besides the Zeeman effect contribution to the order parameter, there is also a contribution from the finite flux that is inserted along the in-plane, or $\hat{y}$, direction of the flake. This magnetic flux from $B_y$ results in a phase modulation encircling the entire circumference of the TI, which is analogous to the Aharonov-Bohm effect[23] and is given as $\frac{\pi B_y x t}{\Phi_0}$. There are thus two contributions to the order parameter: the ordinary Zeeman effect and the Aharonov-Bohm effect, which we call the Zeeman modulation effect (ZME) and flux modulation effect (FME), respectively.

By summing up the three relevant contributions to the phase—the out-of-plane magnetic field $B_z$, the ZME, and the FME—we get the total phase difference between the two junctions:

$$\phi_1(x_1) - \phi_2(x_2) = \frac{2\pi B_z d(x_1+x_2)}{2\Phi_0} + \frac{2B_y(x_1-x_2)g\mu}{\hbar v_f} + \frac{\pi B_y(x_1-x_2)t}{\Phi_0}. \qquad (3)$$

Here, $\phi_1(x_1)$ and $\phi_2(x_2)$ are the phases of the order parameters of superconductor 1 and 2 at the coordinates $x_1$ and $x_2$ respectively along the width of the junction. Based on the phase difference between the two leads, we can model the total transport current along the $\hat{y}$-direction in the Josephson junction using semiclassical methods[17,24]. This analysis is equivalent to summing up all possible quasi-classical trajectories of electron transport, so the total transport current is

$$I(\phi, B_y, B_z) = \int_{-\frac{W_1}{2}}^{\frac{W_1}{2}} \int_{-\frac{W_2}{2}}^{\frac{W_2}{2}} dx_1 dx_2 \frac{1}{d^2+(x_1-x_2)^2} \sin(\Delta\phi + \phi_1(x_1) - \phi_2(x_2)), \qquad (4)$$

where $W_{1(2)}$ is the width of the superconducting lead 1(2), $\Delta\phi$ is the overall phase difference between the superconductors, and $d$ is the distance between the two superconductors. Using Eq. (4), we can calculate the critical current $I_c(B_y, B_z) = \max_\phi I(\phi, B_y, B_z)$ as a function of $B_z$ and $B_y$ to derive the evolution of the Fraunhofer pattern.

Fig. 3b shows simulations of the Fraunhofer pattern for various values of $B_y$ calculated using Eq. (3) and illustrates how the intensity of superconductivity is transferred from the central peak out to the side peaks as $B_y$ is increased. This transfer of intensity is proportional to the momentum shift of the Cooper pair. In terms of the differential resistance as a function of $B_z$ and $B_y$, the transfer of the superconducting intensity can be seen as the formation of two side branches (formed by evolving side peaks) in the differential resistance map with slope $m$. These side branches are visible in the simulation for a symmetric Josephson junction in Fig. 3c and in the data for device 1 (Fig. 2b). The agreement between the simulation and the experimentally observed pattern indicates that the formation of the side branches are a result of the transfer of superconducting intensity due to the additional phase modulation generated by $B_y$. This feature is known to be a key signature of finite momentum pairing and distinguishes the system from typical BCS superconductivity[16,25,26].

Additionally, the slope $m$ of the side branches reflects the relative contributions of finite momentum pairing due to ZME and FME as a function of $B_y$. In the simulations, the slope is defined by a line between the origin and the $n^{\text{th}}$ side lobe as $n$ becomes large (green dashed line in Fig. 3c). By calculating the integral in Eq. (4), the slope of the side branches is estimated as

$$m = \frac{\Delta B_y}{\Delta B_z} = \frac{\pi d/\Phi_0}{\frac{2g\mu}{\hbar v_f} + \frac{2\pi t}{\Phi_0}}. \qquad (5)$$

In Eq. (5), the first and the second terms in the denominator are the contributions of the ZME and the FME to the slope, respectively. Because of the inverse relation, larger slopes reflect smaller ZME and FME contributions. Looking at the FME and ZME contributions separately, we can see that the FME contribution is proportional to the thickness of the flake $t$ since the flux through $B_y$ will increase as the thickness increases. The ZME, on the other hand, is proportional to the intrinsic material parameters of the TI, $\frac{g}{v_f}$, rather than on an external parameter, such as the thickness.

We examine the slope dependence on TI thickness across multiple samples, where devices 2-4 are shown in Fig. 4a-c, respectively, with approximate minima marked and the 2D differential resistance maps rotated so that the lobe minima are vertical for ease of comparison. The slope $m$ for each device is extracted from the minima (see supplementary material) and is illustrated for device 2 by the dashed line in Fig. 4a. To compare the experimental data with theory, we also calculate $m$ for each device based on $t$ and $d$ using Eq. (5). Fig. 4d shows the dependence of $m$ (normalized by an effective $d$ that takes into account flux focusing effects) on thickness, where $m$ is extracted from the data (black) and calculated using theoretical predictions for finite momentum pairing due to ZME alone (red), FME alone (blue), and ZME and FME together (purple).

It is clear that the dominant contribution to the finite momentum shift comes from the FME. In the thickest sample, in particular, the FME closely predicts the slope in the experiment since more flux is enclosed in the flake and the orbital effect therefore has a more significant contribution. However, we find that, the FME is not enough to predict the observed slope from the experiment on its own even when the error bars of slope calculation are taken into account. In fact, the ZME contribution needs to be considered for the theory to more closely match the data, which means that the finite momentum pairing due to the shifted Dirac cones is non-negligible. Therefore, our data is generally better explained by the coexistence of both FME and the unconventional ZME, which is more closely related to the FFLO state.

**Effect of Josephson junction asymmetries on the evolution of the Fraunhofer pattern**

In addition to simulating the Fraunhofer evolution as $B_y$ is applied to an ideal junction, we also consider the effect of asymmetries in the junction geometry and on the evolution of the Fraunhofer pattern. Due to the fact that typical sample fabrication can result in imperfect device configurations and flux focusing effects, it is important to understand what happens to the Fraunhofer pattern as the devices deviate from the ideal junction. For example, as reported in ref. 29, asymmetric features between positive and negative values of $B_z$ often appear in Fraunhofer patterns and can be attributed to a combination of device dependent factors such as disorder and the microscopic structure of the device. We consider some sources of asymmetries in the device configuration to model the effect of these asymmetries on the transport signal.

One form of geometric asymmetry that arises in a Josephson junction is the asymmetry in the width of the two superconducting leads $W_1$ and $W_2$. To model this effect, we introduce the width asymmetry factor $\alpha$, which is the ratio of the two superconducting lead widths and satisfies $W_1 = \alpha W_2$ in Eq. (4). Fig. 5a-b shows how the Fraunhofer evolution changes as we increase the asymmetry between $W_1$ and $W_2$ by increasing $\alpha$. Because finite $\alpha$ breaks the symmetry of $I(\phi, B_y, B_z)$ upon reversing the sign of $B_z$ in Eq. (4), we find that for increasing $\alpha$, the amplitude of the left and right side branches becomes more asymmetric. Another form of asymmetry, as quantified by flux asymmetry factor $\beta$, comes from the flux focusing effect[29]. Due to the screening of the magnetic field inside the superconductor, magnetic field $B_y$ may bend and cause contributions to $B_z$. Since we apply large $B_y$ compared to $B_z$, a very small bending of $B_y$ can cause a large tilt in the Fraunhofer pattern (see supplementary material). We model this effect by replacing $B_z$ with $(B_z - \beta B_y)$ in Eq. (3), which generates the tilt seen in Fig. 5c-d.

After understanding possible origins for anomalous features in the evolution of the Fraunhofer pattern, we can now compare our experimental data with simulations that take into account asymmetry factors. The results are shown in Fig. 6. (See supplementary material for the detailed numerical methods.) The width asymmetry factor $\alpha$ is extracted from scanning electron microscope images of the devices (summarized in Table 1). Due to the difficulty of quantifying the magnitude of the flux asymmetry, we add in an artificial $\beta$ factor to simulations that generates a tilt that best matches with the data for better comparison. As discussed earlier, the overall structure and in-plane field dependence of the Fraunhofer pattern is determined by the ZME, the FME, and the geometry of the junction. However, we find that by also incorporating the sample width asymmetries into the finite momentum pairing model and adding an artificial tilt in the data to take into account flux focusing effects, the theoretical prediction and the experiment agree very well.

**Conclusion**

In conclusion, we observe an anomalous Fraunhofer pattern that is indicative of the presence of finite momentum pairing in 3D topological insulator Josephson junctions that are subjected to in-plane magnetic fields. We identify the two microscopic origins of the finite momentum pairing to be the ZME and the FME. By comparing the slope of the side branches in the anomalous Fraunhofer pattern with the theoretical predictions, we conclude that the measured slope can only be explained by the coexistence of both the ZME and the FME. In particular, the ZME—the contribution associated with the FFLO phase—becomes significant when there is less phase accumulation due to enclosed flux, which occurs for thinner samples. We believe that this is a promising start for the hunt for unconventional superconductivity in a proximity-coupled 3D TI in the presence of in-plane fields, but further work can be done to mitigate the effect of the FME. For example, besides finding thinner flakes, the ZME can be enhanced by increasing $g/v_f$, which can be done by tuning the TI flake closer to the Dirac point. Furthermore, as demonstrated by our work and others, a continued understanding of the effect of non-ideal junctions is conducive to identifying anomalous asymmetric signatures, like the Fraunhofer asymmetry across $B_z$, in transport signals.

**Methods**

3D TI flakes were mechanically exfoliated from bulk $Bi_2Se_3$ crystals onto $Si/SiO_2$ substrates, and thicknesses were identified using atomic force microscopy. After identifying suitable flakes, superconducting electrodes were defined using standard ebeam lithography techniques. 40-65 nm of NbTi/NbTiN was then sputtered onto the device using an rf source following a brief ion mill to clean off the surface. Devices were then wire-bonded and measured in a dry dilution unit that reaches a base temperature of T = 25 mK and has a three-axis vector magnet.

**References**

[1] Bardeen, J., Cooper, L. N. & Schrieffer, J. R. Theory of superconductivity. *Phys. Rev.* **108**, 1175–1204 (1957).
[2] Fulde, P. & Ferrell, R. A. Superconductivity in a strong spin-exchange field. *Phys. Rev.* **135**, A550–A564 (1964).

3. Larkin, A. I. & Ovchinnikov, Y. N. Inhomogeneous state of superconductors. *Sov. Phys. JETP* **20**, 762–769 (1965).
4. Radovan, H. A. *et al.* Magnetic enhancement of superconductivity from electron spin domains. *Nature* **425**, 51-55 (2003).
5. Bianchi, A., Movshovich, R., Capan, C., Pagliuso, P. G., & Sarrao, J. L. Possible Fulde-Ferrell-Larkin-Ovchinnikov Superconducting State in CeCoIn$_5$. *Phys. Rev. Lett.* **91**, 187004 (2003).
6. Singleton, J. *et al.* Observation of the Fulde-Ferrell-Larkin-Ovchinnikov state in the quasi-two-dimensional organic superconductor $\kappa$-(BEDT-TTF)$_2$Cu(NCS)$_2$(BEDT-TTF=bis(ethylene-dithio)tetrathiafulvalene). *J. Phys.: Condens. Matter* **12**, L641 (2000).
7. Lortz, R. *et al.* Calorimetric Evidence for a Fulde-Ferrell-Larkin-Ovchinnikov Superconducting State in the Layered Organic Superconductor $\kappa$-(BEDT-TTF)$_2$Cu(NCS)$_2$. *Phys. Rev. Lett.* **99**, 187002 (2007).
8. Mayaffre, H. *et al.* Evidence of Andreev bound states as a hallmark of the FFLO phase in κ-(BEDT-TTF)2Cu(NCS)2. *Nat. Phys.* **10**, 928–932 (2014).
9. Kenzelmann, M. *et al.* Coupled superconducting and magnetic order in CeCoIn5. *Science* **321**, 1652–1654 (2008).
10. Y. Matsuda, and H. Shimahara. Fulde–Ferrell–Larkin–Ovchinnikov State in Heavy Fermion Superconductors. J. Phys. Soc. Jpn. **76**, 051005 (2007).
11. R. Beyer and J. Wosnitza Low Temp. Phys. **39**, 225 (2013).
12. Young, B. L. *et al*. Microscopic Evidence for Field-Induced Magnetism in CeCoIn$_5$. *Phys. Rev. Lett.* **98**, 036402 (2007).
13. Z. Zheng, et al. FFLO Superfluids in 2D Spin-Orbit Coupled Fermi Gases. *Scientific Reports* **4**, 6535 (2014).
14. Dimitrova, O. & Feigel'man, M. W. Theory of a two-dimensional superconductor with broken inversion symmetry. *Phys. Rev. B* **76**, 014522 (2007).
15. Yoshida, T, Sigrist, M, & Yanase, Y. Pair-density wave states through spin-orbit coupling in multilayer superconductors. *Phys. Rev. B* **86**, 134514 (2012).
16. Park, M. J. *et al.* Fulde-Ferrell states in inverse proximity-coupled magnetically doped topological heterostructures. *Phys. Rev. B* **96**, 064518 (2017).
17. Hart, S. *et al.* Controlled finite momentum pairing and spatially varying order parameter in proximitized HgTe quantum wells. *Nat. Phys.* **13**, 87-93 (2017).
18. Veldhorst, M. *et al.* Josephson supercurrent through a topological insulator surface state. *Nat. Mater.* **11**, 417–421 (2012).
19. Cho, S. *et al*. Symmetry protected Josephson supercurrents in three-dimensional topological insulators. *Nat. Commun.* **4**, 1689 (2013).
20. Lee, J. H. *et al.* Local and nonlocal Fraunhofer-like pattern from an edge-stepped topological surface Josephson current distribution *Nano Lett.* **14** 5029-34 (2014).
21. Tinkham, M. *Introduction to Superconductivity* (Dover Publications Inc., 2013).
22. Zhang, H. *et al.* Topological insulators in Bi$_2$Se$_3$, Bi$_2$Te$_3$ and Sb$_2$Te$_3$ with a single Dirac cone on the surface. *Nat. Phys.* **5**, 438-442 (2009).
23. Aharonov, Y & Bohm, D. Significance of Electromagnetic Potentials in the Quantum Theory. *Phys. Rev.* **115**, 485 (1959).
24. Mohammadkhani, G., Zareyan, M., & Blanter, Ya. M. Magnetic interference pattern in planar SNS Josephson junctions, Phys. Rev. B 77, 014520 (2008).

[25] Kim, Y, Park, M. J., & Gilbert, M. J. Probing unconventional superconductivity in inversion-symmetric doped Weyl semimetal *Phys. Rev. B* **93**, 214511 (2016).

[26] K. Yang and D. F. Agterberg Josephson Effect in Fulde-Ferrell-Larkin-Ovchinnikov Superconductors *Phys. Rev. Lett.* **84**, 4970 (2017).

[27] Liu, C. -X. *et al.* Model Hamiltonian for topological insulators. *Phys. Rev. B* **82,** 045122 (2010).

[28] Wolos, A. *et al.* g-factors of conduction electrons and holes in Bi2Se3 three-dimensional topological insulator *Phys. Rev. B* **93**, 155114 (2016).

[29] Suominen, H. J. *et al.* Anomalous Fraunhofer interference in epitaxial superconductor-semiconductor Josephson junctions. *Phys. Rev. B* **95**, 035307 (2017).

**Acknowledgements**

N.M. and A.Q.C. acknowledge research support from the National Science Foundation (NSF) under Grant No. DMR 14-11067. M.J.G. and M.J.P. acknowledge support from the NSF under Grant No. CAREER ECCS-1351871. N.M., M.J.G., A.Q.C., and M.J.P. acknowledge support from the NSF under Grant No. DMR 17-10437. G.J.M and Y.X. acknowledge support from the U.S. Department of Energy under Grant No. DE-SC0012368. This work was carried out in part in the Frederick Seitz Materials Research Laboratory Central Facilities, University of Illinois. M.J.P. also acknowledges Mark R. Hirsbrunner for helpful discussions.

**Author contributions**

A.Q.C. fabricated devices and carried out transport measurements. M.J.P. performed transport simulations. S.T.G. helped A.Q.C. with experimental set up. Y.X. and G.J.M. grew the $Bi_2Se_3$ crystals. A.Q.C., M.J.P., M.J.G., and N.M. analyzed the data and wrote the manuscript.

**Figure 1: Measurement configuration and Device 1 Fraunhofer pattern**

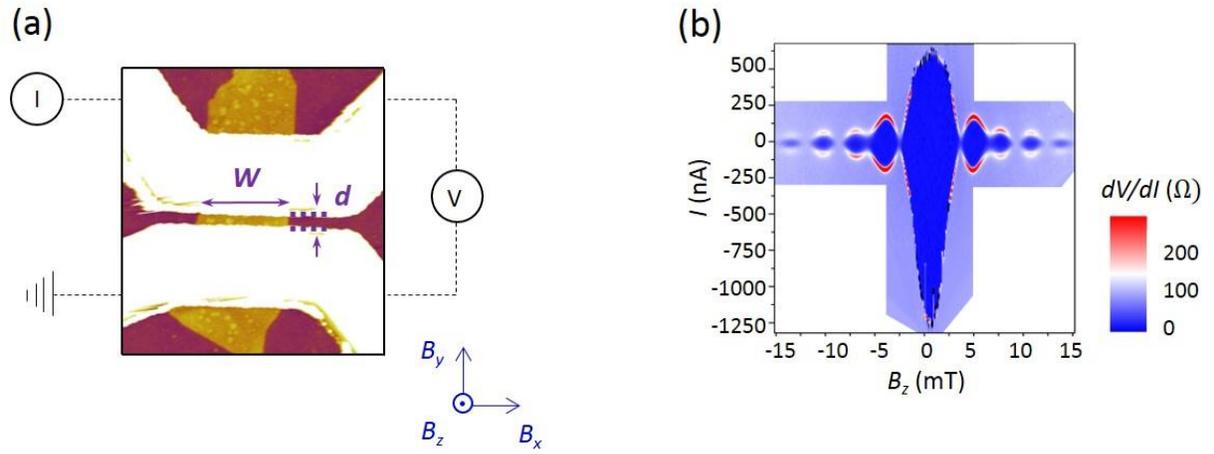

(**a**) AFM image of the S-TI-S Josephson junction: superconducting leads (white) on a $Bi_2Se_3$ flake (yellow) exfoliated onto a substrate (red). Measurement scheme and magnetic field configurations are also shown. (**b**) Conventional Fraunhofer pattern for device 1 ($t \sim 9$ nm, $d \sim 140$ nm, and $W \sim 920$ nm). The conventional pattern has a principal peak at $B_z = 0$.

**Figure 2: Device 1 Fraunhofer evolution**

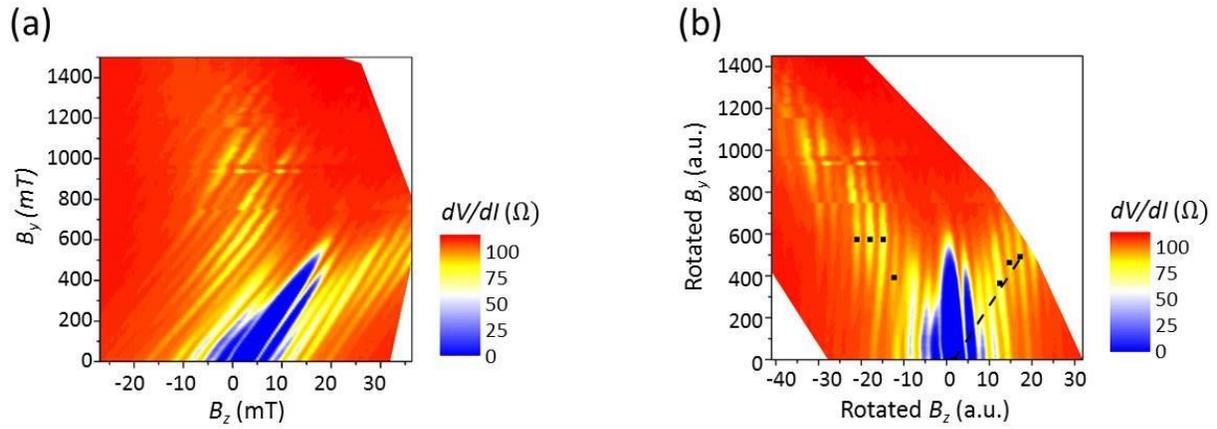

(**a**) Evolution of the Fraunhofer patterns for device 1. (**b**) Fraunhofer evolution for device 1 that has been rotated so that the lobe minima are vertical, making it easier to compare across samples. There is a side branch feature that develops as $B_y$ is applied to the junction, which can be quantified as a line with slope *m* (dashed line). Black dots mark approximate locations of minimum resistance at different side lobes as a guide to the eye.

**Figure 3: Finite momentum shift and simulation of trident pattern for ideal junction**

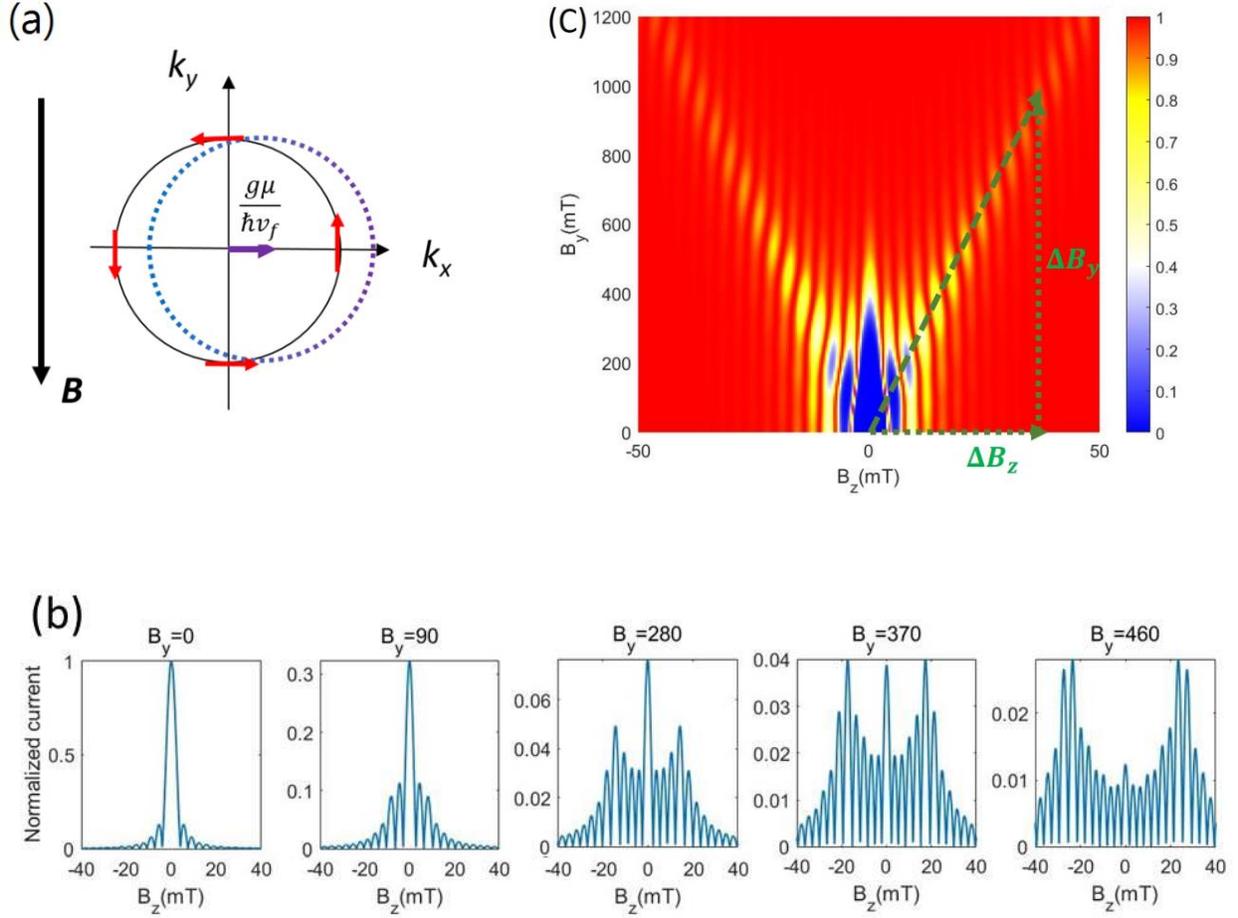

(**a**) The shifted Fermi surface and spin texture (red arrows) of the TI due to a finite Zeeman effect. To form a spin singlet, Cooper pairs acquire a nonzero center of momentum, as indicated by the purple arrow. (**b**) Simulations of the Fraunhofer pattern for different values of $B_y$. We find that the intensity of superconductivity is transferred outwards to higher values of $B_z$ as $B_y$ is increased. (**c**) Evolution of the Fraunhofer pattern for a symmetric Josephson junction due to finite momentum pairing. The differential resistance is calculated and normalized to 1. The slope of the side branch is indicated by the dashed green line.

**Figure 4: Fraunhofer evolution for devices with thicker flakes**

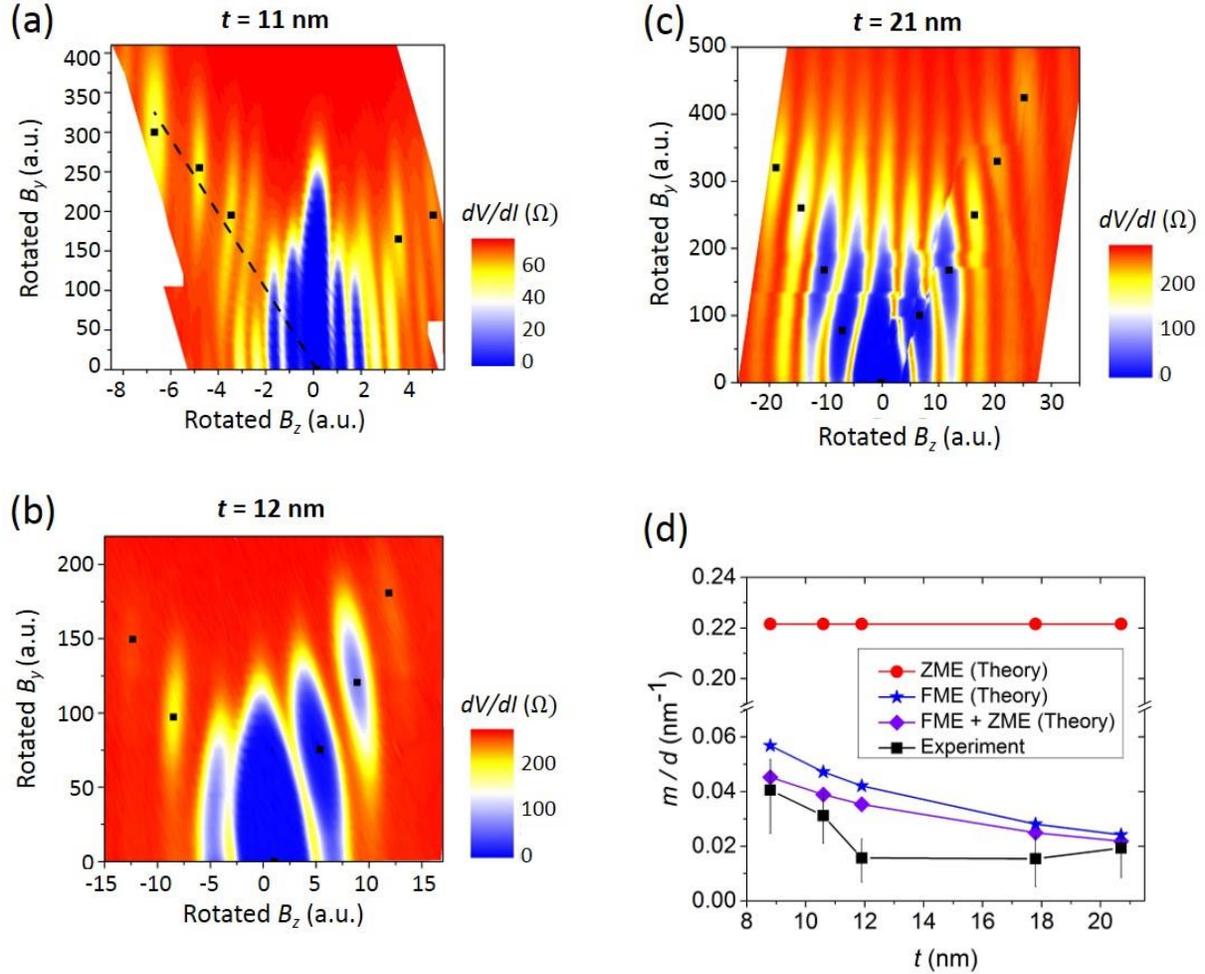

(**a**)-(**c**) Fraunhofer evolution for devices 2-4, which have different flake thicknesses. For comparison, the data was rotated so that the lobe minima are vertical for better comparison and approximate minima are marked with black dots. The side branch slope is illustrated for device 2 in (a). (**d**) The relation between the slope of the side branch *m* (normalized by effective electrode distance *d*) and thickness *t* of the TI flake. Experimental data (with error bars for deviations in extracted slopes) is compared with simulations for each device using a finite momentum pair model. The theory and data matches best for a model that takes into account both ZME and FME. $v_f = 5 \times 10^5 m/s$, $g = 19$ are used in the simulations[27,28].

**Figure 5: Simulation of effect of junction asymmetry and field inhomogeneity**

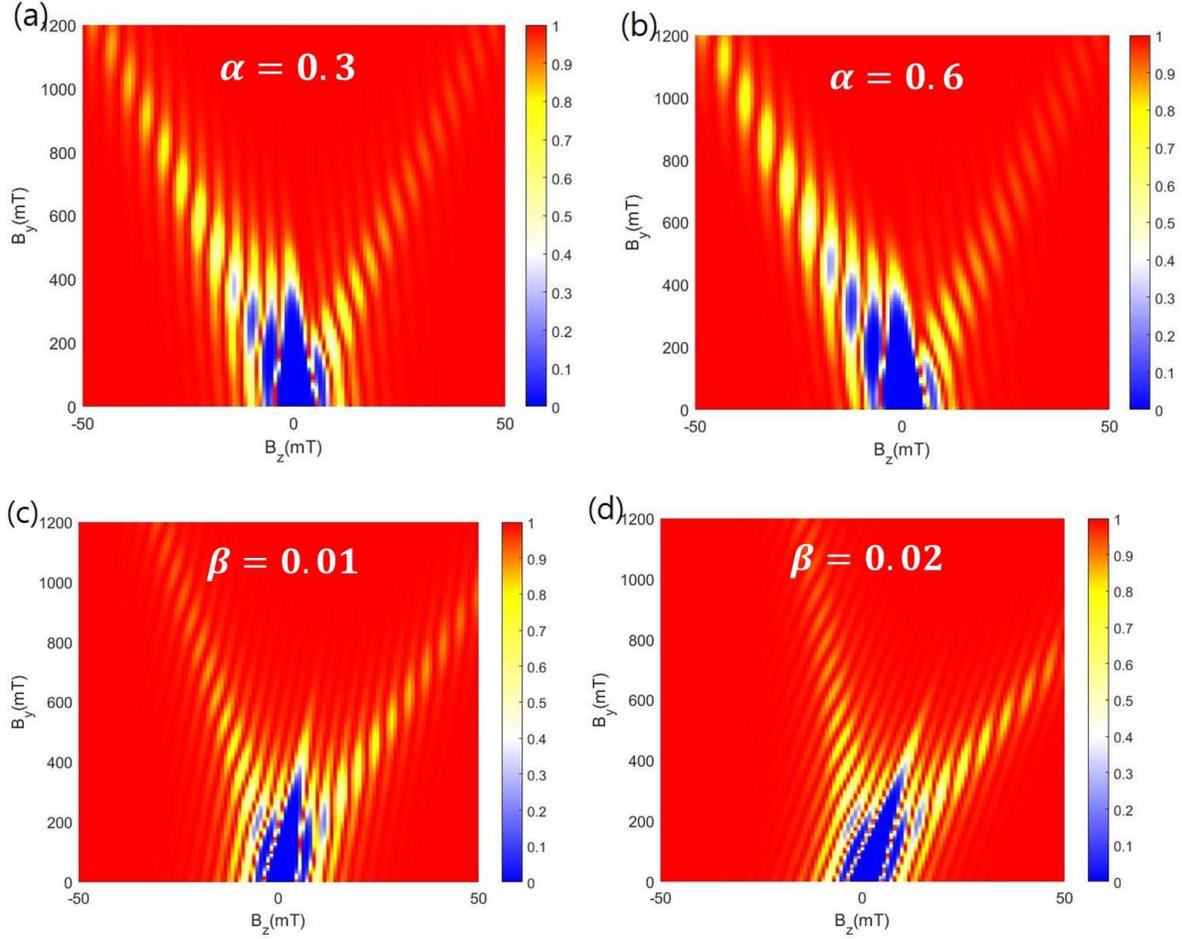

The width and field asymmetry dependence of the Fraunhofer pattern. (**a**)-(**b**) When a width asymmetry factor $\alpha$ is added to the model, we find that the signal becomes asymmetric between positive and negative $B_z$. Here, the amplitude of the left side lobes increases relative to the amplitude of the right side lobes. We used the values $\alpha = 0.3, 0.6$ respectively. (**c**)-(**d**) When the asymmetry factor $\beta$ is introduced to the model, the Fraunhofer patterns are shifted along $B_z$, which can be seen as a tilt introduced to the 2D differential resistance map. We used $\beta = 0.01, 0.02$ respectively.

**Figure 6: Evolution of the Fraunhofer pattern across different devices**

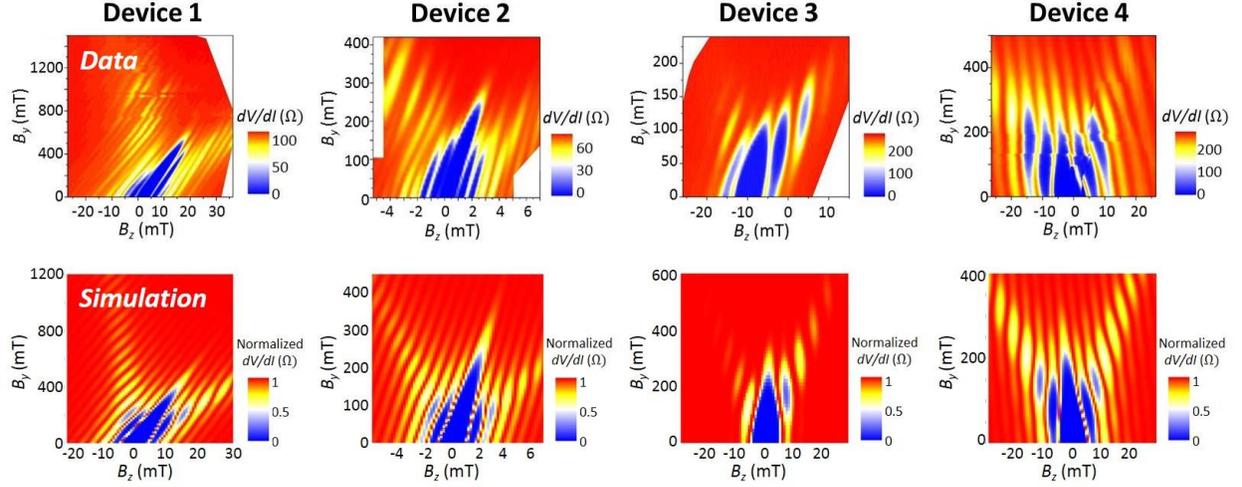

Comparison of the evolution of the Fraunhofer pattern when $B_y$ is applied to devices 1-4. The experimental data (top) agrees well with the normalized differential resistance of the simulations (bottom) that are based on a finite momentum shift model and take into account width asymmetries delineated in Table 1. An additional tilt is added to the simulations for better comparison with the data. (A possible origin for the tilt is a flux asymmetry, which is difficult to quantify.) In all the devices, side branches with slope $m$ appear as an in-plane field $B_y$ is added. The appearance of these side branches is indicative of interference in the phase modulation due to $B_z$ and $B_y$.

**Table 1: Dimensions for devices 1-5**

| Device number | $t$ (nm) | Average $W$ (nm) | $d$ (nm) | $\alpha = \dfrac{W_1}{W_2}$ |
|---|---|---|---|---|
| 1 | 9 | 860 | 140 | 1.07 |
| 2 | 11 | 1930 | 240 | 1.04 |
| 3 | 12 | 570 | 160 | 1.15 |
| 4 | 21 | 500 | 270 | 1.00 |
| 5 | 18 | 940 | 220 | 1.04 |

**Supplementary material**

### A. Effective electrode spacing $d$

For a conventional Fraunhofer pattern, the nodes occur at $B_z = \frac{n\Phi_0}{A}$, where magnetic flux quantum $\Phi_0 = \frac{h}{2e}$, $A$ is the area of the junction, and $n$ is an integer[1]. However, as discussed in Ref. 2, there can be variable node spacing due to a field-dependent flux focusing from the superconducting leads.

Though the overall features of our zero in-plane Fraunhofer patterns follow a conventional pattern in that they have a large central peak at $B_z = 0$ and decaying side peaks, the location of the nodes of our device Fraunhofer patterns do deviate from $B_z = \frac{n\Phi_0}{A}$ if we were to naively use area measurements obtained from an SEM image. We assume that this deviation in node position is due to flux focusing, which reduces the distance through which flux actually penetrates. The effective distance $d$ can then be extracted from the value of the first Fraunhofer node: $d = \frac{B_{z,\text{first node}}}{W}$, where $W$ is the width of the electrode as measured in an SEM image.

### B. Discussion of slope extraction from the experimental data

As discussed in the main text, the slope of the side branches $m$ characterizes the transfer of superconductivity intensity from the central Fraunhofer peak out to higher values of $B_z$ as $B_y$ is increased. In the simulations, this is defined by a line that connects the origin and the $n^{\text{th}}$ side lobe as $n$ becomes large. Ideally, a slope could be similarly extracted from the data by connecting a line from the origin to the minimum of the side lobe furthest from the origin. However, the range of our experimental data is usually limited so that the minima of only 2 to 5 side lobes are lifted from the $B_y = 0$ axis. Furthermore, the experimental data can have asymmetric features (like the ones discussed in the main text or other anomalous features) that can cause the location of individual side lobes to deviate from the side branch line drawn from the origin to a large $n^{\text{th}}$ side lobe.

Here, we discuss how we calculate the slope of the side branches to best approximate a characteristic $m$ for each device. After the locations of the minima of the lifted side lobes are extracted from the differential resistance map (which has been rotated so that lobe minima are vertical), a line can be drawn from the origin to each of the minima, as shown for device 1 in Fig.

S1a. Fig. S1b shows the slope values $|m_n|$ (normalized by an effective electrode spacing $d$) for devices 1-5. Each color represents the slope of a line drawn to a different lifted side lobe, where R(L) denotes a side lobe to the right (left) of $B_z = 0$ and the number corresponds to the $n^{th}$ lifted side lobe. Because $|m_n|/d$ do not deviate substantially from each other for each device and all have a similar dependence on the flake thickness $t$, we take the average of all the $|m_n|/d$ for each device to be a good approximation to observe the relationship between slope and flake thickness. This average slope $m$ is shown in Fig. S1a (dashed green line) and is the slope value used in the main text.

### C. Details of the numerical methods

In this section, we provide a detailed description of the numerical simulations used in the main text. After we calculate the critical current using Eq. (4), we numerically generate Fraunhofer patterns as a function of $B_y$ for each device. We find that a non-zero $B_y$ transfers the intensity of superconductivity from $B_z = 0$ out to higher values of $B_z$. To illustrate this feature, we present various slices of the Fraunhofer pattern as a function of $B_y$ in Fig. S2. Once we derive a set of Fraunhofer patterns as a function of $B_y$, we can draw the surface contour plot of the critical current as a function of both $B_y$ and $B_z$. In this plot, the transfer of intensity appears as side branches.

To compare with the experimental data, we map the critical current map to a differential resistance map. In order to transform the Fraunhofer pattern into the differential resistance map, we model an effective thermal noise using Ambegaokar-Halperin (AH) theory, which is given as

$$V = \frac{4\pi}{\gamma}\left\{(e^{\pi\gamma x} - 1)^{-1}\left[\int_0^{2\pi} d\theta f(\theta)\right]\left[\int_0^{2\pi} d\theta \frac{1}{f(\theta)}\right] + \left[\int_0^{2\pi}\int_\theta^{2\pi} d\theta d\theta' \frac{f(\theta)}{f(\theta')}\right]\right\}^{-1} . \quad (A.1)$$

Here $V$ and $x$ are normalized voltage and current, respectively. $f(\theta) = e^{\frac{1}{2}\gamma(x\theta + \cos\theta)}$. $\gamma$ is an effective dimensionless parameter representing thermal fluctuations, which is typically given as $\gamma = \frac{\hbar I_C}{eK_bT}$. By fitting $dV/dI$ calculated using Eq. (A.1) to the measured I-V characteristics of the experiment, we extract the effective value of the thermal fluctuations. We find that the extracted temperature from the AH theory does not necessarily match the base temperature of the experiment (T = 25 mK), indicating a possible increase in the temperature inside the device or fluctuations due to noise, but we also point out that it has been reported that the AH theory has shown quantitative discrepancies in underdamped junctions. Therefore, we instead treat $\gamma$ as a fitting parameter to account for the effective fluctuations in the device.

Fig. S3 shows the comparison between the critical current map of the evolution of the Fraunhofer pattern and the corresponding differential resistance map. When the critical current is larger than the excitation current, the differential resistance becomes negligible, corresponding to the flat blue region. If the excitation current is higher than the critical current, it is seen as red resistive region in the differential resistance map. Table S1 summarizes the numerical parameters used to generate Fig. 6 in the main text.

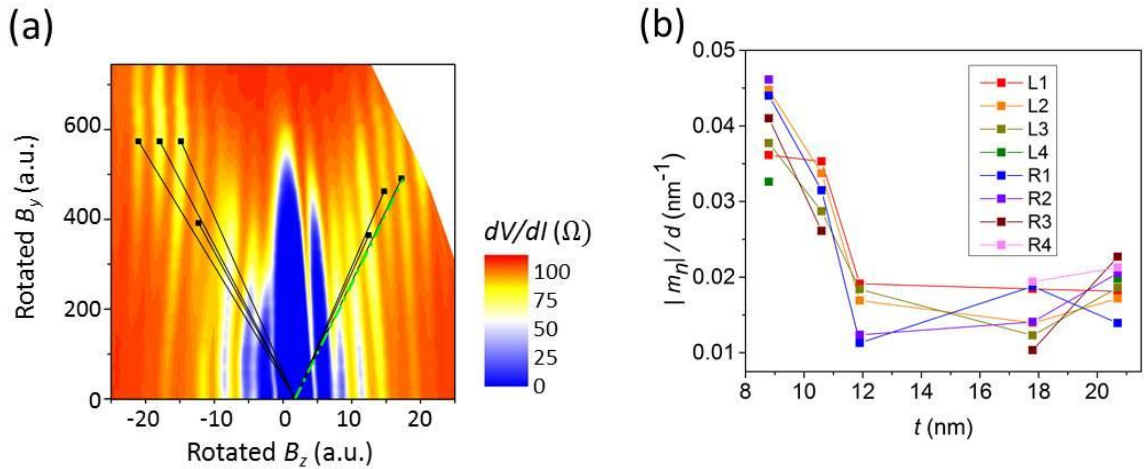

Fig. S1 (a) Lines to the minima of each lifted side lobe have slope $|m_n|$ and are shown (black lines) for device 1. The average of these slopes, $m$, is also shown (dashed green line) and corresponds to the slope value used in the main text. (b) Normalized slopes $|m_n|$ are extracted for each device and plotted as a function of thickness $t$. Each color represents the slope to a different side lobe.

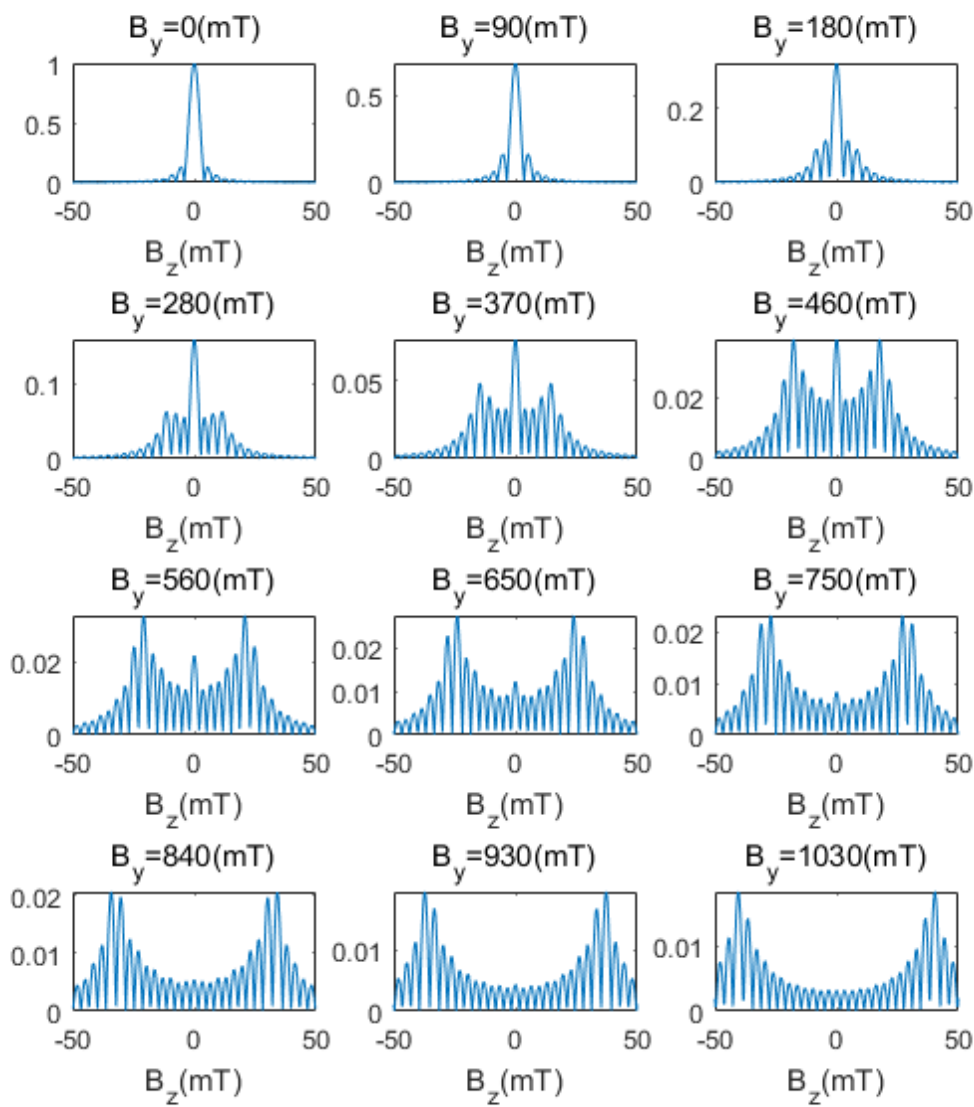

Fig. S2 Slices of the Fraunhofer pattern for a fixed $B_y$. We find that the intensity of the superconductivity is transferred from the center to the sides as we increase $B_y$.

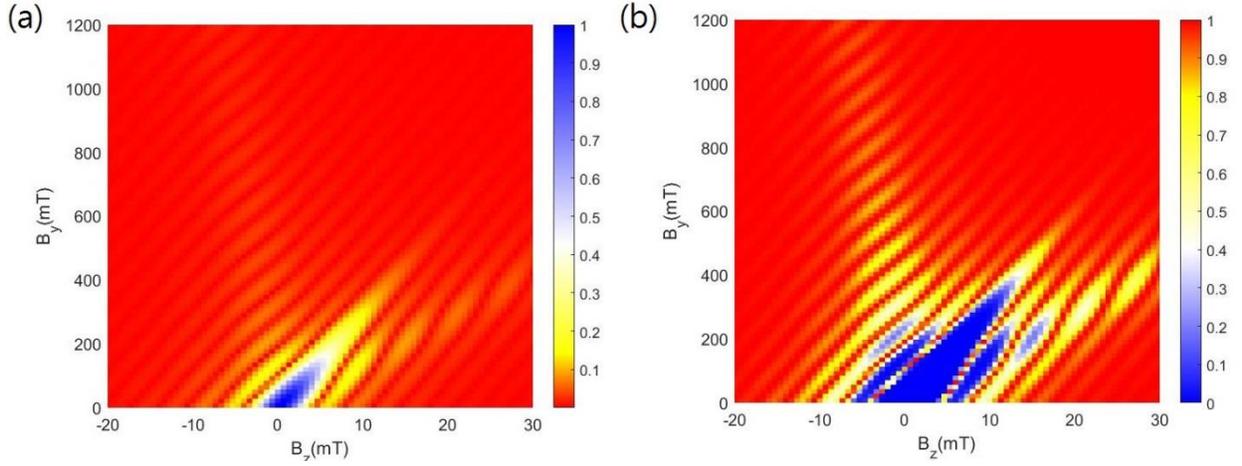

Fig. S3 (a) Numerically calculated critical current map of device 1. (b) The corresponding normalized differential resistance calculated using Eq. (A.1). If the excitation current is lower than the critical current, the differential resistance map shows a flat zero resistance region(blue).

Table S1: Numerical parameter used in Fig. 6 in the main text.

| Device number | $\alpha = \dfrac{W_1}{W_2}$ | $\beta$ | $\gamma(B=0)$ |
|---|---|---|---|
| 1 | 1.07 | 1/40 | 58 |
| 2 | 1.04 | -1/200 | 58 |
| 3 | 1.15 | 0 | 38 |
| 4 | 1.02 | 1/100 | 58 |
| 5 | 1.00 |  |  |

References

[1] Tinkham, M. *Introduction to Superconductivity* (Dover Publications Inc., 2013).

[2] Suominen, H. J. et al. Anomalous Fraunhofer interference in epitaxial superconductor-semiconductor Josephson junctions. Phys. Rev. B 95, 035307 (2017).